\documentclass[aps,pra,twocolumn,10pt,showpacs]{revtex4-1}
\usepackage{graphicx}
\usepackage{bm}
\usepackage{amsthm,amsfonts,amstext,amsmath,amssymb}
\usepackage{latexsym}
\usepackage[margin,index]{fixme}
\usepackage[utf8]{inputenc}
\usepackage[makeroom]{cancel}
\usepackage[T1]{fontenc}
\usepackage[usenames,dvipsnames]{color}

\newcommand{\lr}[1]{\left( #1 \right)}

\newcommand{\be}{\begin{equation}}
\newcommand{\ee}{\end{equation}}
\newcommand{\bea}{\begin{eqnarray}}
\newcommand{\eea}{\end{eqnarray}}

\def\ie{{i.e.},\ }

\def\cf{{cf}.\ }

\newcommand*\Laplace{\mathop{}\!\mathbin\bigtriangleup}

\begin{document}
	\title{Quantum Hall Physics with Cold Atoms in Cylindrical Optical Lattices}

	\author{Mateusz~\L\k{a}cki${}^{1,2,3}$, Hannes~Pichler${}^{1}$, Antoine~Sterdyniak${}^{2}$, Andreas~Lyras${}^{4}$, Vassilis~E.~Lembessis${}^{4}$, Omar Al-Dossary${}^{4,5},$ Jan~Carl~Budich${}^{2}$, Peter~Zoller${}^{1,2}$} 
	\affiliation{
		\mbox{${}^{(1)}$ Institute for Quantum Optics and Quantum Information of the Austrian Academy of Sciences, A-6020 Innsbruck, Austria;}
		\mbox{${}^{(2)}$ Institute for Theoretical Physics, University of Innsbruck, A-6020 Innsbruck, Austria;}		
		\mbox{${}^{(3)}$ Instytut Fizyki imienia Mariana Smoluchowskiego, Uniwersytet Jagiello\'nski, \L{}ojasiewicza 11, 30-348 Krak\'ow, Poland;}
		\mbox{${}^{(4)}$ Department of Physics and Astronomy, College of Science, King Saud University, Riyadh 11451, Saudi Arabia;}		
		\mbox{${}^{(5)}$ The National Center for Applied Physics, KACST, P.O. Box 6086, Riyadh 11442, Saudi Arabia.}		
	}		
	\date{\today}
	\begin{abstract} We propose and study various realizations of a Hofstadter-Hubbard model on a cylinder geometry with fermionic cold atoms in optical lattices. The cylindrical optical lattice is created by copropagating Laguerre-Gauss beams, i.e.~light beams carrying orbital angular momentum. By strong focusing of the light beams we create a real space optical lattice in the form of rings, which are offset in energy. A second set of Laguerre-Gauss beams then induces a Raman-hopping between these rings, imprinting phases corresponding to a synthetic magnetic field (artificial gauge field). In addition, by rotating the lattice potential, we achieve a slowly varying flux through the hole of the cylinder, which allows us to probe the Hall response of the system as a realization of Laughlin's thought experiment. We study how in the presence of interactions fractional quantum Hall physics could be observed in this setup.  \end{abstract}
\maketitle

\begin{figure}[ht]
	\includegraphics[width=\columnwidth]{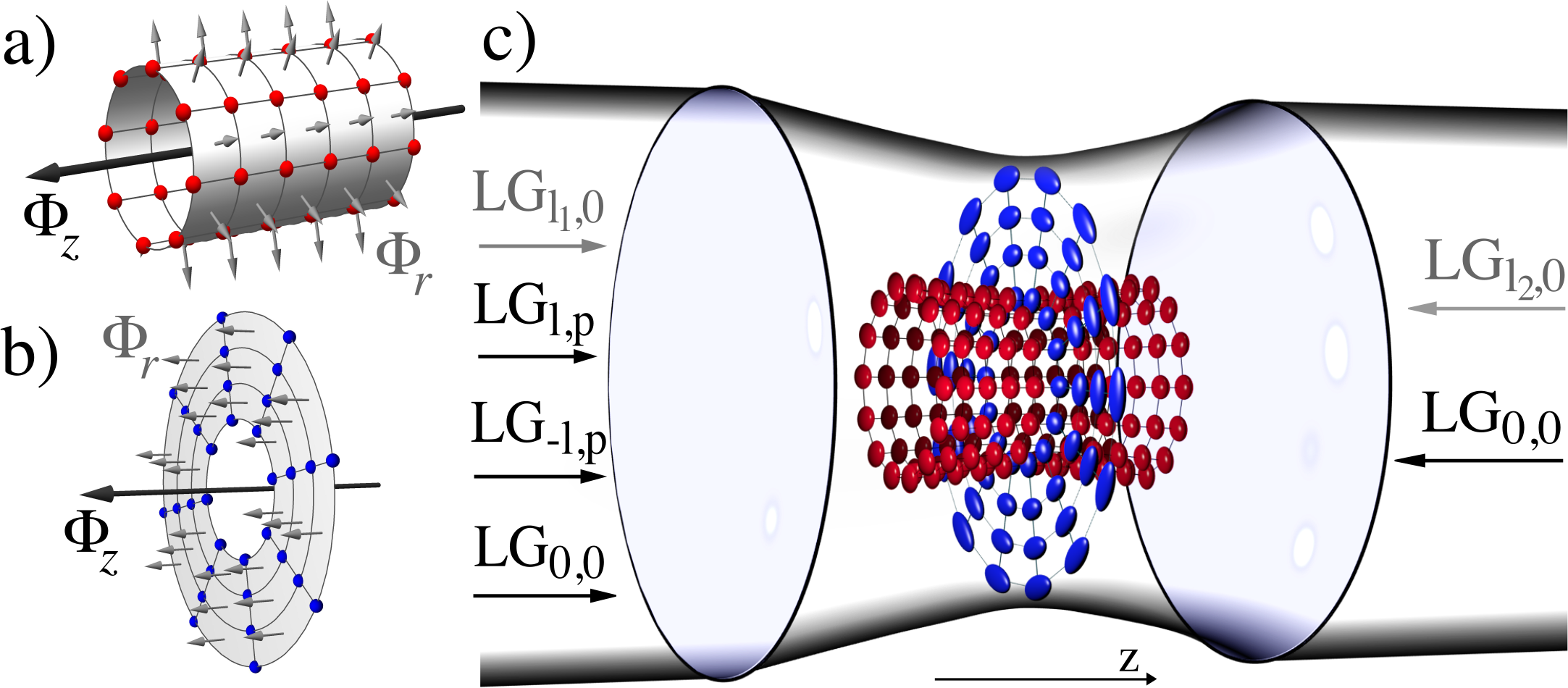}
	\caption{(Color online) Lattices with real cylindric (red sites) and ring shaped (blue sites) geometry around the focal plane of a lens system. Artificial magnetic fluxes through the surface ($\Phi_r$) and the interior ($\Phi_z$) [see panels a) and b)] are feasible in both scenarios. Panel c) outlines the proposed setup. A Gaussian standing wave creating a lattice along $\hat{z}$ direction (black $LG_{0,0}$). Tightly focused lasers carrying orbital angular momentum $l$ and $-l$, respectively (black $LG_{l,p}$  and $LG_{-l,p}$), create the azimuthal lattice potential. Counterpropagating lasers with orbital angular momenta $l_1$  and $l_2$  (gray $LG_{l_1,0}$ and $LG_{l_2,0}$) create $\Phi_r$ via Raman processes.}
	\label{fig:one}
\end{figure}

{\emph{Introduction. --}}
For our understanding of quantum many-body systems, considering spherical, cylindrical, toroidal, or even more exotic geometries has in many situations proven to be of key importance \cite{Fradkin2013}. This is because phenomenologically distinct physical properties may be revealed by imposing various boundary conditions. Prominent examples along these lines include persistent currents, protected edge states \cite{Halperin1982}, topological ground state degeneracies \cite{Wen2007}, and spectral flow in response to fluxes threading the holes of a system \cite{Laughlin1981}. The natural question, as to what extent such theoretically intriguing constructions can become experimentally viable, can be seen as a challenge in quantum engineering. A photonic crystal with M\"obius strip geometry \cite{Simon2015} and an artificial flux threading an atomic ring potential \cite{Campbell2014} have been recently realized experimentally.
For cold atoms in optical lattices \cite{Bloch2008,Lewenstein2012} with planar geometries, remarkable progress has been reported in devising \cite{Jaksch2003,Duan2006,GerbierDalibard,Cooper2011,Cooper2013,DalibardReview,GoldmanReview} and experimentally probing \cite{Aidelsburger2011,Sengstock,Ketterle2013,Aidelsburger2013,Atala2014,Aidelsburger2015} gauge fields and topological states \cite{HasanKane2010,QiReview2011}. It is the purpose of the present work to present a microscopic model for a {\emph{cylindrical optical lattice}} in real space that realizes the Hofstadter model~\cite{Hofstadter1976} of fermionic atoms subject to a perpendicular synthetic magnetic field [see Fig.~\ref{fig:one}a)]. Our proposal builds on, and is motivated by advances in generating light beams carrying orbital angular momentum (OAM) \cite{Padget2008,Zeilinger2012}.  The simulation of artificial magnetic fields on non-trivial geometries has also been addressed with the experimentally intriguing  idea of a {\em synthetic dimension} \cite{Celi2014,Lewenstein2015,Cooper2015,Spielman2015,Fazio2015,Marie2015}, i.e.~a manifold of internal states coupled by Raman lasers that are interpreted as (a small number of) lattice sites. In contrast, our work focuses on  {\em real space} cylindrical lattices, which becomes essential once interactions are included to realize fractional Quantum Hall (FQH) physics  \cite{Tsui1982,Laughlin1983,Prange1990}.

For non-interacting atoms we present a microscopic realization of the fermionic Hofstadter Hamiltonian ~\cite{Hofstadter1976} on a cylinder
\be
H_0=-\!\sum\limits_{\mathbf{j}} J^z_{\mathbf{j}} e^{i \varphi_{\mathbf{j}}} a_{\mathbf{j}}^\dag a_{\mathbf{j}+\hat z} - \!\sum\limits_{\mathbf{j}} J_{{\mathbf{j}}}^\phi a_{\mathbf{j}}^\dag a_{\mathbf{j}+\hat \phi_{\mathbf{j}}} +\textrm{h.c.},
\label{eqn:totalHamiltonian}
\ee
where $J^z_\mathbf{j}$ and $J_{\mathbf{j}}^\phi$ denote amplitudes for nearest neighbor hopping to lattice site $\mathbf{j}$ by a displacement vector $\hat z$ and $\hat \phi_{\mathbf{j}}$ in axial and azimuthal direction, respectively, $a_\mathbf{j}$ annihilates a spinless fermion at site $\mathbf{j}$, and the spatially dependent phase factors $ e^{i \varphi_{\mathbf{j}}}$ affect the neutral atoms analogous to the effect of a magnetic field on charged particles. Adding interactions, the total Hubbard Hamiltonian is given by $H=H_{0}+H_{I}$. For spinless fermionic atoms the natural (minimal)  interaction is the nearest-neighbor interaction $H_{I} = U\sum_{\langle \mathbf{i},\mathbf{j}\rangle} n_{\mathbf{i}}n_{\mathbf{j}}$, where $n_{\mathbf{i}} = a_{\mathbf{i}}^\dag a_{\mathbf{i}}$ is the particle number on site $\mathbf{i}$.

Below, we address two main questions. First, we detail how a cylindrical optical lattice with a synthetic magnetic flux $\Phi_r$ piercing its surface can be created. This is achieved by employing tightly focused laser beams that carry OAM [see Fig.~1c) for a schematic], both to create the lattice potential and to engineer an artificial magnetic field via Raman-assisted tunnelling. In addition, by rotating the cylindrical lattice potential around its axis, we are capable of mimicking a slowly varying flux $\Phi_z$ threading the hole of the cylinder in the axial direction [see Fig.~\ref{fig:one}a)] (see also Ref.~\cite{Campbell2014}). We argue how the resulting spectral flow hallmarking the quantum Hall effect can be experimentally observed. These ingredients allow us to implement Laughlin's original thought experiment \cite{Laughlin1981}, explaining the integer quantum Hall effect \cite{Klitzing1980,Laughlin1981,TKNN1982} in terms of spectral flow.

Second, going beyond the single particle picture, we investigate numerically the effect of local interactions and outline how FQH physics \cite{Tsui1982,Laughlin1983,Prange1990} can occur in our cylindrical model. We compare the Hofstadter-Hubbard model on a real space cylinder proposed here to its counterpart involving a synthetic dimension \cite{Cooper2015,Fazio2015}. Including interactions, synthetic and real dimensions are shown to lead to a qualitatively different behavior. The physical reason behind this is that spatial locality of many-body interactions and localized wave functions giving rise to topological protection occur naturally in real samples  but may be non-generic or require fine tuning in the context of synthetic dimensions.\\

\begin{figure}
 	\includegraphics[width=1\columnwidth]{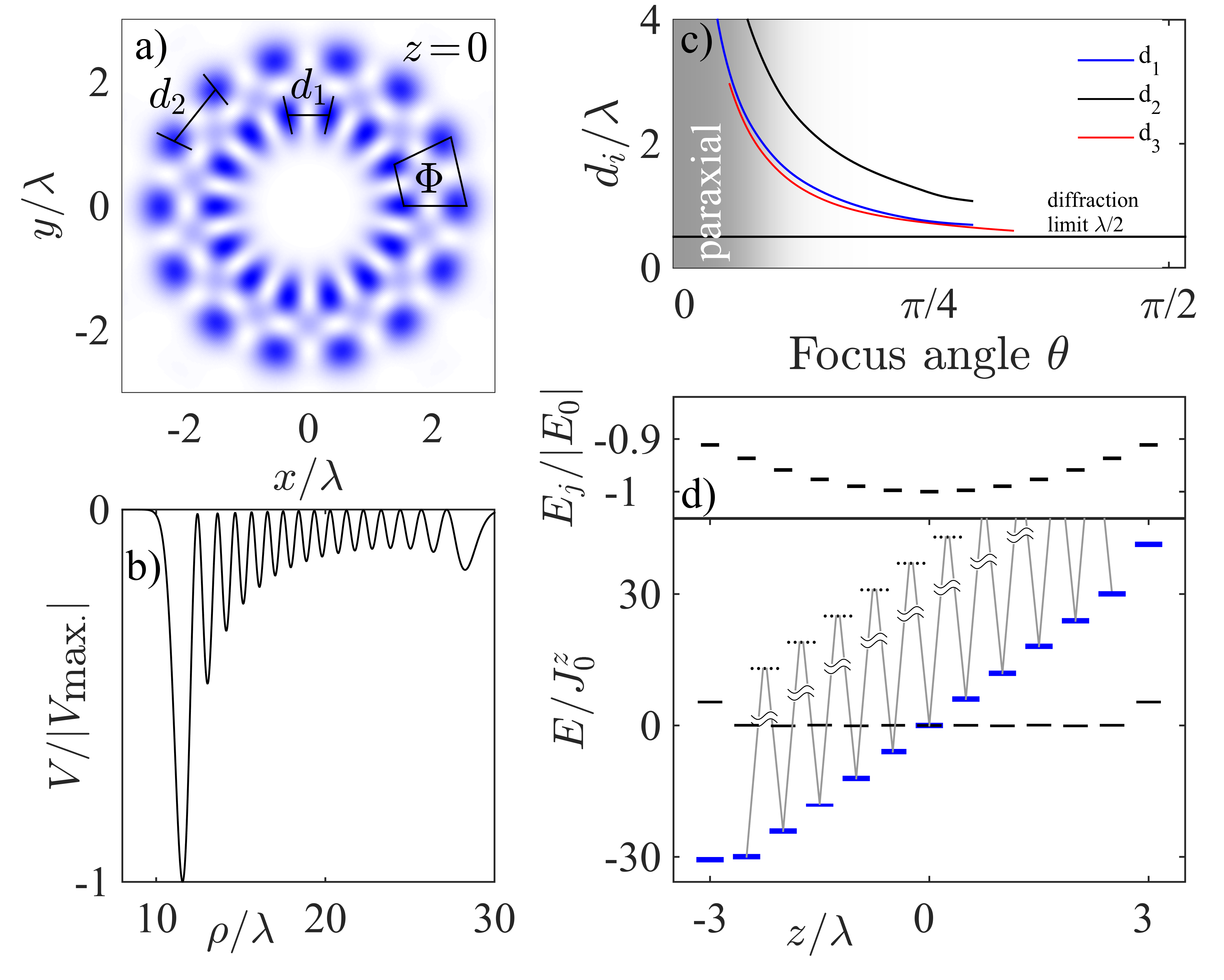}
 	\caption{(Color online) Panel a): Total intensity pattern of two interfering beams: $LG_{7,1}$ and $LG_{-7,1}$ including nonparaxial effects, realizing a small two-ring system in the focal plane. Panel b): radial cut through the optical potential for an azimuthal lattice (interfering $LG_{60,15}$ and $LG_{60,-15}$). Panel c): Intersite distances $d_1, d_2$ [see a)] converging to the diffraction limit for strong focusing, away from the grey paraxial regime. $d_3$ refers to the innermost ring in our main setup in [for parameters see Fig. \ref{fig:fig3}]. The upper part of d) shows the natural variance of single particle energy levels due to strong focusing. In the lower part it is compensated for by a flattening potential (black levels, for details see \cite{supplementary}), and then tilted by a magnetic field gradient (blue levels show an exemplary tilt) to make two-photon Raman processes between neighboring rings resonant. The dotted lines indicate the virtual intermediate atomic states involved in the Raman process. }	
 	\label{fig:fig2}
 \end{figure}

{\emph{Cylindrical optical lattices. --}} 
We are interested in creating optical lattices with annulus or cylinder geometry as depicted in Figs~\ref{fig:one}a) and~~\ref{fig:one}b). Moreover, building on Raman assisted tunneling techniques \cite{Jaksch2003,GerbierDalibard,DalibardReview} and rotation of the lattice potential, respectively, we aim at engineering the two fluxes $\Phi_r$ and $\Phi_z$. The spinless fermionic atoms are assumed to occupy a single Zeeman $m$-state, where the quantization axis is defined by homogeneous magnetic field $\vec B= B_0 \hat z.$
The key ingredients for our proposal are tightly focused light beams carrying OAM \cite{Allen2003,Andrews2011,Andrews2012}. A~large focus angle is essential to reach lattice spacings $d$ on the order of the optical wavelength $\lambda\equiv 2\pi/k$ and thus sufficiently large energy scales $
\frac{\lambda^2}{4d^2}E_R$, where $E_R=\frac{\hbar^2k^2}{2m}$ is the laser recoil energy. Some basic features, however, can be understood in terms of a paraxial description valid for weakly focused light. There, Laguerre Gaussian (LG) laser beams  $LG_{l,p}$ with electric field ${\mathbf{E}}_{l,p}(\rho,\phi,z) \sim \mathbf{e}_\sigma {\xi}^{|l|} L_p^{|l|}(\xi^2)e^{-\xi^2/2}e^{il\phi+ikz}$ describe light carrying an OAM of  $l\hbar $ per photon \cite{Allen1992}. Here $\mathbf{e}_\sigma$ is the polarization and $L_p^l$ denote associated Laguerre polynomials and  $\xi=\sqrt{2}\frac{\rho}{w_0}$. In this framework, an intensity modulation $I\sim \cos^2(l\phi)$ around the beam axis arises naturally when two beams  $LG_{l,p}$ and $LG_{-l,p}$ are interfered \cite{Amico2005,FA2007,Cominotti2014}. Together with the radial dependence of the intensity, it gives rise to lattice potentials consisting of $p+1$ concentric rings with $2l$ azimuthal sites each [see Fig.~\ref{fig:fig2}a)]. Beams with $l=25$ and $p=10$ have been realized experimentally \cite{Thirugnanasambandam2010,Senatsky2012,Zupancic2013,Ngcobo2013,Preiss2015}. The paraxial approximation would limit the lattice constant $d$ to values much larger than $\lambda/2$ \cite{Nemoto1990}. This limitation stands in stark contrast to the key requirements of reaching experimentally viable energy scales. Here we investigate {\emph{tightly focused}} beams with OAM, thus going conceptually beyond the simplified picture of the paraxial approximation. The dependence of $d$ on the focusing angle $\theta$ \cite{Andrews2012} is displayed in Fig.~\ref{fig:fig2}c), showing that $d\approx \lambda/2$ is achievable in the non-paraxial regime.

To achieve strong focussing, we consider two aplanatic lenses with focal length $f$ to focus incident LG beams propagating along the $z$-axis, i.e.~the optical axis of our setup [see Fig.~\ref{fig:one}c)].
For an incident LG beam with angular momentum $l$ and circular polarization  ${\bf e}_+=-({\bf e}_x+i{\bf e}_y)/\sqrt{2},$ the electric field  ${\bf E}_{l,p}$ close to the focal plane is given by \cite{Monteiro2009,Novotny2012}
\bea
{\bf E}_{l,p}= e^{il\phi}\bigg({\cal E}_{|l|}^{+}{\bf e}_++{\cal E}_{|l+1|}^{z}e^{i\phi}{\bf e}_z-{\cal E}_{|l+2|}^{-}e^{2i\phi}{\bf e}_-\bigg).
\label{eqn:E}
\label{eqn:Iintegrals}
\eea
We note that in this non-paraxial regime the polarization can no longer be separated from the spatial mode profile. Moreover, the focused field is no longer transverse and the different polarization amplitudes are given by ${\cal E}_{l}^{\sigma}(\rho,z)=E_0\int_0^{\theta_{\rm m}}\!\! d\theta \sin\theta\,g_l(\theta)\sqrt{\cos{\theta}} h_\sigma(\theta)J_l(k\rho\sin\theta) e^{ikz\cos\theta}$, where we abbreviated $h_{\pm}(\theta)=1\pm\cos\theta$ and $h_z(\theta)=-\sqrt{2}i\sin\theta$, as well as $g_l(\theta)=\xi^{|l|}L_p^{|l|}(\xi^2)e^{-\xi^2/2}$, with $\xi=\sqrt{2}\frac{f}{w_0}\sin\theta$. The parameters $E_0$ and $w_0$ specify the intensity and the waist of the beams incident on the lens. The numerical aperture of the lens enters via $NA=\sin \theta_m$. The Bessel functions  are denoted by $J_l$. 

By superimposing two such laser beams with the same propagation direction, but with opposite $l$, we obtain an intensity pattern $I_\sigma(\rho,\phi, z)$, which is invariant under rotation of $\pi/l$ around the optical axis [see Fig.\ref{fig:fig2}a)] in each polarization component $I_\sigma$. Importantly, even though the azimuthal phase dependence of the electric field is not simply $e^{il\phi}$ [\cf Eq.~\eqref{eqn:Iintegrals}] like in the paraxial treatment, this symmetry is guaranteed due to the circular polarization of the incident light. For details we refer to the supplemental material. The optical potential is then given by $V(\vec r)=\sum_{\sigma\in\{+,-,\pi\}} \alpha_\sigma I_\sigma (\rho,\phi,z)$, where $\alpha_\sigma$ is the atomic polarizability, and we assume $\alpha_{\sigma}\equiv \alpha<0$ below. By adding two Gaussian beams in a standing wave configuration we create a lattice along the optical axis. 

In this setting we can realize two different geometries: 

One possibility is an {\em annulus geometry} realized by confining atoms to the focal plane  [Fig.~\ref{fig:one}b), blue sites in Fig.~\ref{fig:one}c)], consisting of a~series of concentric ring lattices with $d\sim \lambda/2.$ In the radial direction, the rings are decoupled by an energy offset stemming from the radially varying laser intensity [see Fig. \ref{fig:fig2}b)].  Such an offset can be used to engineer the synthetic gauge field $\Phi_r$ by coupling the rings with Raman lasers \cite{Ketterle2013,Aidelsburger2013}. 

Instead we focus here on the {\em cylinder geometry}, obtained by restricting atoms to the innermost (energetically lowest) ring of each disk [see Fig. \ref{fig:one}c)], and coupling the rings by Raman-assisted tunnelling \cite{Jaksch2003, GerbierDalibard, DalibardReview}. 
We note that strong focussing leads to an inhomogeneity along the $z$-direction of the cylindrical lattice potential, where the dominant effect stems form the decrease of laser intensity away from the focal plane. This leads to the trapping potential [see Fig. \ref{fig:fig2}d)] but also to a minor $z$-dependence of the azimuthal hopping \cite{supplementary}. We observe that the latter dependence is even smaller for high radial modes $p>0$.   

  \begin{figure}
  	\includegraphics[width=\columnwidth]{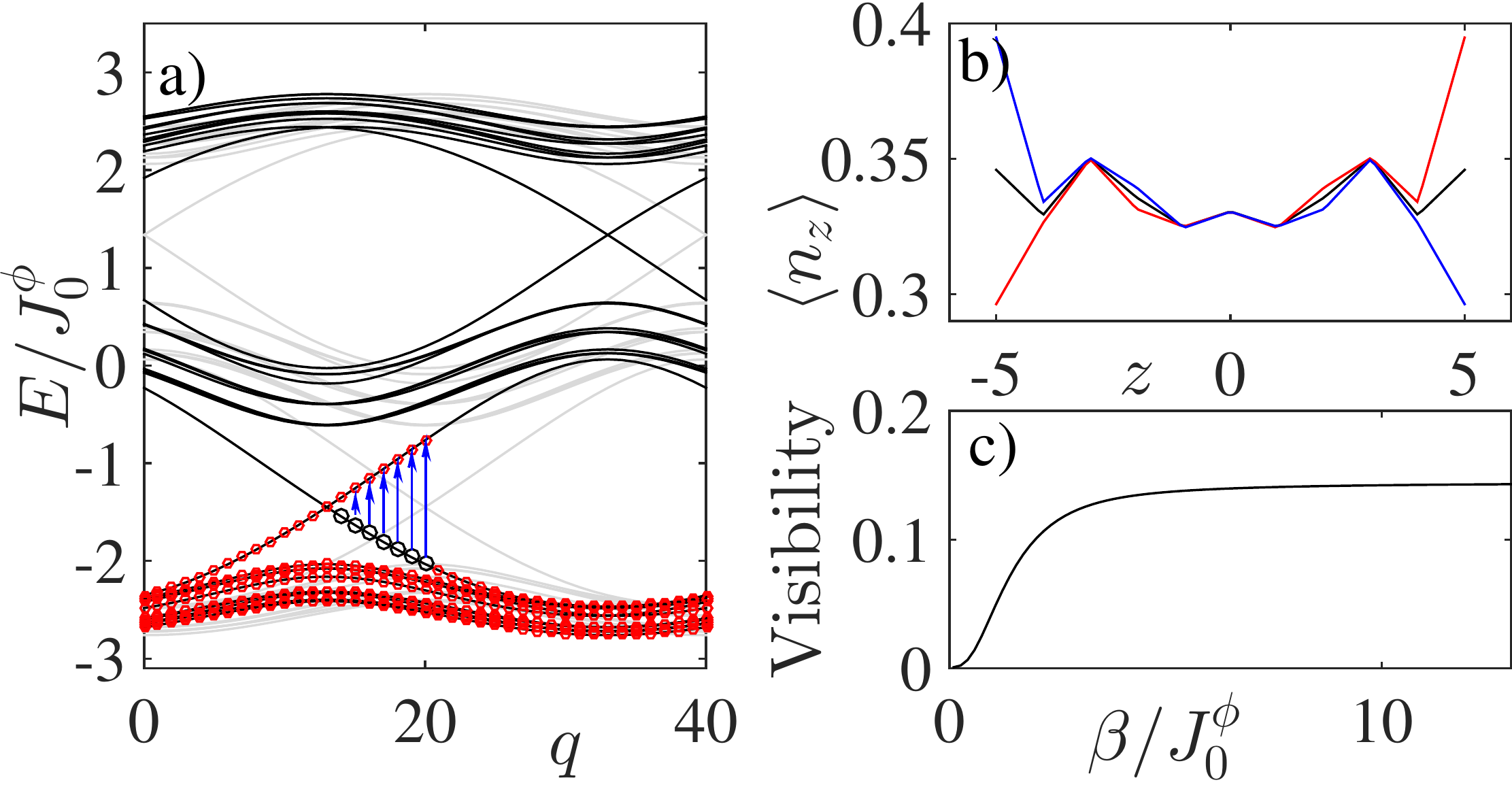}
  	\caption{ (Color online) Laughlin experiment on a cylinder. Panel (a): Single particle energy spectrum of (\ref{eqn:totalHamiltonian}) with a threaded flux $\Phi_{z}=14\pi$ (black lines) and $\Phi_{z}=0$ (gray lines) as a function of the dimensionless azimuthal lattice momentum $q$ which is quantized in units of $2\pi\hbar/N_m$,  where $N_m=40$ is the number of magnetic unit cells. Red circles show states occupied after flux threading, while blue arrows indicate population transfer with respect to the initial state population (black circles).   Panel (b) shows $\langle n_z^\Phi  \rangle$ --- the average density of the gas after flux $\Phi$  has been threaded (black --- $\Phi_{z,0}=0,$ red --- $\Phi_{z,f}=14\pi$, blue --- $\Phi_{z,f}=-14\pi$). Panel (c): Temperature dependence of the visibility $v=(\langle n_5^{\Phi=\Phi_{z,f}} \rangle - \langle n_5^{\Phi=\Phi_{z,f}}  \rangle ) / \langle n_5^{\Phi=\Phi_{z,0}} \rangle.$ Parameters are: maximal azimuthal lattice height $V_\phi=8.5E_R,$ axial --- $V_z=7E_R$. The lattice consists of 11 rings each containing 120 sites ($l=60, p=15, l_1=20, l_2=-20$). }
  	\label{fig:fig3}
  \end{figure}

{\it Artificial magnetic field on a cylinder}. -- To mimic the magnetic flux $\Phi_r$, we here adapt the Raman assisted tunnelling scheme to our cylindrical optical lattice. This scheme consists of two steps. First, the optical lattice potential is tilted in one spatial direction, for example by a magnetic field gradient, which renders the hopping in this direction off-resonant. Second, we restore hopping via a two-photon Raman process. Using LG beams ($LG_{l_1,0}$ and $LG_{l_2,0}$, see Fig. \ref{fig:one}c)) for this Raman process allows us to imprint the required phase pattern for the flux $\Phi_r$.

In our setup, strong focusing introduces naturally an optical potential tilted in the axial direction [see Fig.~\ref{fig:fig2}d)]. However, this tilt is not a linear staircase, requiring in principle separate Raman transition frequencies to couple neighbouring rings, which would practically limit this scheme to a few coupled rings.
Cylinders with a large number of rings can be achieved by compensating for the nonlinear energy offsets by adding an appropriate spatially-dependent AC-Stark shift, as familiar from flattening of inhomogeneous optical lattices \cite{Schneider2015}. By adding a magnetic field gradient \cite{Ketterle2013,Aidelsburger2013} in the $\hat{z}$ direction, we achieve a linear tilting as illustrated in Fig. \ref{fig:fig2}d).  This allows us not only to couple the rings by a single pair of Raman beams generating a homogeneous flux $\Phi_r$, but residual nonlinearities away from the flattened region also provide a sharp cutoff for the cylinder in the axial direction [see Fig. \ref{fig:fig2}d)].
 
For a  Raman pair of lasers carrying OAM $l_1$ and $l_2$, the hopping amplitude between sites $\mathbf{j}$ and $\mathbf{j}+\hat{z}$ acquires the phase dependence 
\begin{align}
{\cal J}_{\mathbf{j}}^z=J^z_{\mathbf{j}}e^{i(l_2-l_1) \phi_\mathbf{j}}=J^z_{\mathbf{j}}e^{i\varphi_{\mathbf{j}}},\label{eq:Syntetic_gauge}
\end{align}
where  $J_\mathbf{j}^z\in\mathbb{R}$ and $\phi_\mathbf{j}$ is the azimuthal angle of the lattice site $\mathbf{j}$.
An atom hopping around a plaquette of the lattice with $2l$ azimuthal sites now picks up a nonzero phase $\delta \phi=2s \pi /l$, ($s\equiv l_2-l_1$). 
Due to strong focusing, different polarization components carry different phase dependencies [see Eq. \eqref{eqn:Iintegrals}]. However, these non-paraxial effects do not change the phase pattern \eqref{eq:Syntetic_gauge} imprinted on the hopping elements.
The final tight-binding model is then of the form of the Hofstadter Hamiltonian on a cylinder (\ref{eqn:totalHamiltonian}). Below we explore its band structure including the adiabatic Hall response, which may be probed by spectroscopic techniques and time-of flight imaging, and the role of interactions.

{\it Flux threading and Hall response}. -
The Hall response of a cylindrical system can be probed along the lines of Laughlin's famous thought experiment \cite{Laughlin1981}, where a small electric field in azimuthal direction is imposed on a gas of electrons by adiabatically threading a~flux in axial direction through the cylinder. A quantum Hall system such as our Hofstadter model (\ref{eqn:totalHamiltonian}) then responds with a current in axial direction that is generated by spectral flow. The integer number of particles transported between the ends of the cylinder per threaded flux quantum then equals the quantized Hall conductance in units of $e^2/h$.

Our present setup offers a natural way to realize such a scenario.
An artificial time dependent flux may be implemented by a slowly accelerated rotation with frequency $\Omega(t)$ of the cylindrical lattice around its symmetry axis, which translates the spectrum in quasimomentum by $\Delta k(t) =  2m \Omega(t) r^2/\hbar$ close to the focal plane, where $r$ denotes the radius of the cylinder. Experimentally, such a rotation is readily achieved by a frequency detuning of the counter-propagating laser beams generating the lattice potential. At the level of the tight-binding model (\ref{eqn:totalHamiltonian}), it imprints a complex phase to the azimuthal hopping, \ie $J_{\mathbf{j}}^\phi\to J_{\mathbf{j}}^\phi e^{i\Delta k(t)}$. The small detunings caused by the rotation do not affect the other building blocks of our proposal. In Fig.~\ref{fig:fig3}b), we show the Hall response to such an axial flux of a system of ultracold fermions with magnetic flux $\Phi_r=2\pi/3$ per plaquette [see Eq.~(\ref{eqn:totalHamiltonian})] at zero temperature and $1/3$ filling of the lattice, \ie with one filled Chern band [see Fig.~\ref{fig:fig3}a)]. Depending on the orientation of the magnetic flux, atoms are transferred in different directions between the edges of the cylinder  [see Fig.~\ref{fig:fig3}b)]. For finite temperatures, the visibility of this charge pumping is shown in Fig.~\ref{fig:fig3}c).

{\emph{Effect of interactions and fractional filling. --}} We now turn to the role of interactions, focusing on the possibility of realizing fractional quantum Hall (FQH) physics \cite{Tsui1982,Laughlin1983,Prange1990} on a cylinder. 
We are particularly interested in small cylindrical systems that may be realized in a first generation of experiments. Using exact diagonalization, we assess which signatures of FQH physics could be observable.
We focus on one-third filling of the lowest band at flux $\Phi_r=2\pi/3$. In this situation, short-ranged interactions are expected to stabilize FQH states \cite{Kol1993,Lukin2005}, more specifically the $1/3$-Laughlin state \cite{Laughlin1983,Kol1993,Bauer2015}. In our model study we assume a nearest neighbor interaction between the spinless fermions $H_{I} = U\sum_{\langle \mathbf{i},\mathbf{j}\rangle} n_{\mathbf{i}}n_{\mathbf{j}}$, as naturally realized microscopically by magnetic interactions in dipolar Er and Dy fermionic quantum gases \cite{DyEr}.
We compare our results for the {\em real space cylinder} to its counterpart involving a {\em synthetic dimension} \cite{Celi2014}. While the free Hamiltonians are the same and give rise to a metallic phase, the interaction in the presence of a synthetic dimension is naturally of {\emph{infinite range}} \cite{Ye2014,Fallani2014,Fallani2014a} along the synthetic dimension, since different lattice sites are not spatially separated in real space, i.e.
$H_{\mathrm{syn}} = \frac{U}{2} \sum_{i=0}^{2l-1} \left( \sum_{z} n_{i\hat \phi+z\hat z} \right)^2 $.

\begin{figure}[t]
	\includegraphics[width=\columnwidth]{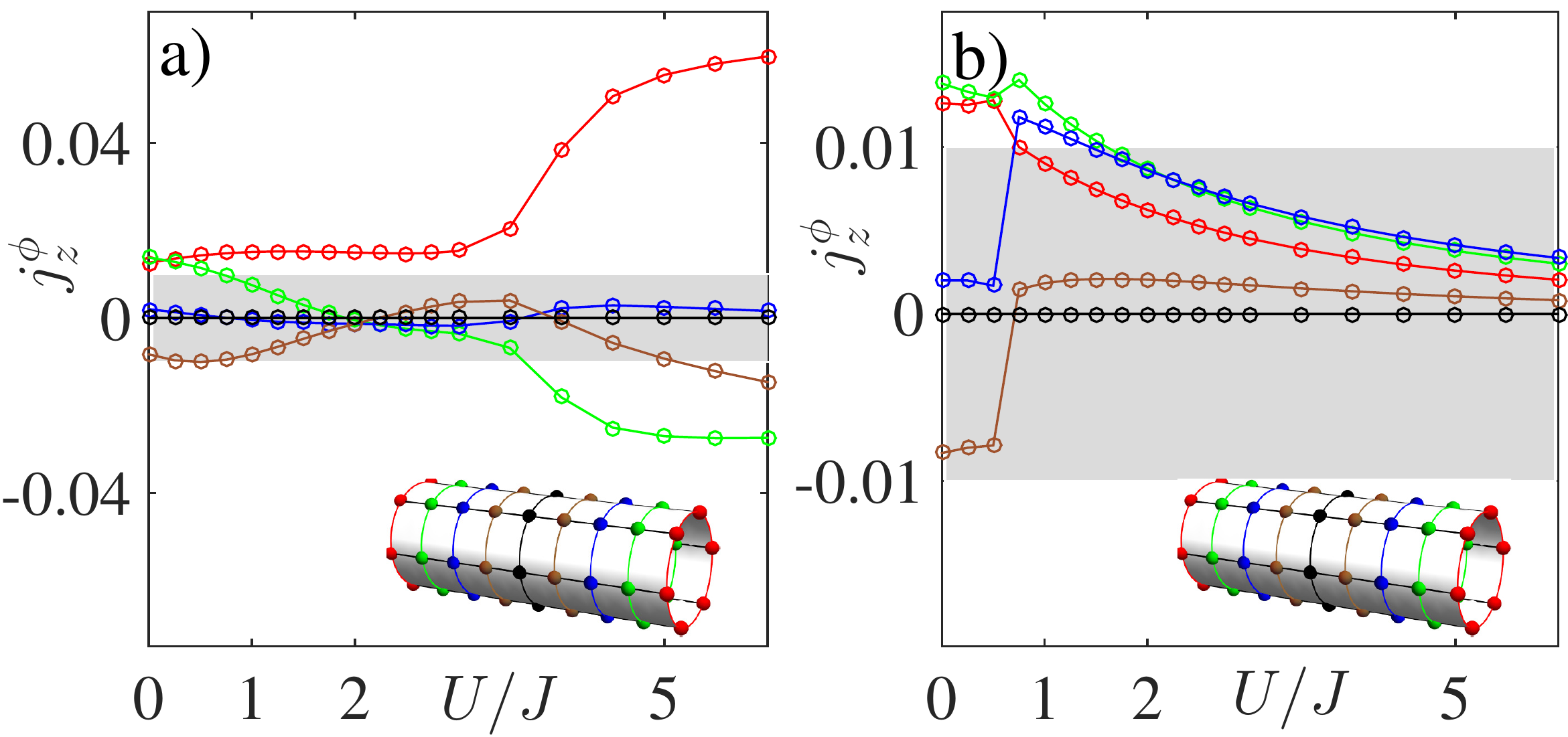}
	\caption{(Color online) Azimuthal current $j^{\phi}_z$ around the rings of the cylinder as a function of the (repulsive) interaction strength for the real cylinder a) and the synthetic dimension analog b). We consider $6$ fermions on a cylindrical optical lattice made of $9$ rings with $6$ sites each (i.e. $l=3, l_1=1,l_2=-1$). Accounting for parity symmetry in $z$-direction, only the $5$ independent rings are shown.}
	\label{fig:fig4}
\end{figure}

In the cylinder geometry, a simple probe for a FQH state is to look at the interaction-induced emergence of chiral edge currents. To this end, we compute the value of the azimuthal current $j^{\phi}_z = \langle i J_{{\mathbf{j}}}^\phi a_{\mathbf{j+\hat{\phi_j}}}^{\dag} a_{\mathbf{j}} + \text{h.c.}\rangle$ in the groundstate on the different rings $z$ as a function of interaction strength. As the system is translation invariant in the azimuthal direction, the current only depends on $z$. Our numerical results are shown in Fig.~\ref{fig:fig4}.
For the real cylinder geometry [see Fig.~\ref{fig:fig4}a)], the groundstate remains in the zero momentum sector as expected for an incompressible liquid such as the Laughlin state. At small interaction, the current of the groundstate with local interactions is of the same order on each ring of the cylinder. With increasing interaction strength, we observe an increasing dominance of the edge current. While, given the small system sizes, the system does not display a fully insulating bulk yet, this may be seen as a first signature of FQH physics that could be observed in experiments on very small cylinders. Experimentally going beyond numerically accessible system sizes, this signature is expected to become increasingly clear. By contrast, the groundstate of the system involving a synthetic dimension [see Fig.~\ref{fig:fig4}b)] has a non-zero momentum for $U \gtrapprox 0.6J$ and displays smaller edge currents, decreasing with the interaction strength.

\emph{Outlook. -- }  The present proposal can be extended to create an optical lattice with torus geometry: this can be achieved with two concentrical cylinders which, in our setup, are naturally decoupled due to a~radial energy shift. Coupling of the cylinders at the edges can be restored  by photon-assisted tunnelling, effectively sewing together the patterns to a seamless torus.

\acknowledgments
We thank M.~Aidelsburger, I.~Bloch, M.~Greiner, N.~Regnault, O.~Romero-Isart, and C.~Schweizer for discussions. PZ thanks the King Saud University, Riyadh, for hospitality during a visit. M\L\  was supported by the Polish National Science Center, project no. 2013/08/T/ST2/00112 and the Foundation for Polish Science (FNP). Work at Innsbruck was supported by SFB FoQuS (FWF Project No.~F4006- N18), the ERC Synergy Grant UQUAM, the Austrian Ministry of Science BMWF as part of the Konjunkturpaket II of the Focal Point Scientific Computing at the University of Innsbruck, and from EU via SIQS.

\bibliographystyle{apsrev}

\vspace{1cm}
\section*{Supplementry material}

In this Supplementary Material, we present a more detailed and quantitative analysis of the optical potential. Moreover, we microscopically determine the relevant interaction parameters for a~realistic cylindrical lattice potential.

\section{Quantitative properties of the potential}
\label{sec:Qpotp}
The interference pattern defining the lattice structure in the azimuthal direction is created by superposing two tightly focused copropagating beams that carry opposite orbital angular momentum (OAM) $l_1=l, l_2=-l.$ 
For the specific example presented in this work we have chosen the Laguerre-Gaussian modes incident on the lensing system to have azimuthal and radial indices $l=60$ and $p=15$ respetively. The focusing of the beam is described by the ratio of the focal length to the paraxial waist of a~pre-lens beam, which we choose to be $f/w_0=10.35$. The asymptotic propagation angle of the outermost intensity maximum determines the focus angle of the whole beam to the value of $\theta=56.5^\circ$ [see equations of the main text]. 

As discussed in the main text, the total intensity $I_{\textrm{tot}}\sim |E_l + E_{-l}|^2$ splits into polarization components: $I=\sum_{\sigma\in\{+,-,\pi\}} I_\sigma (\rho,\phi,z),$ where each of them is of the form 
\be
I_\sigma(\rho,\phi, z)=a_\sigma(\rho,z)+b_\sigma(\rho,z)\cos^2(l\phi + c_\sigma(\rho,z)),
\label{eqn:abc}
\ee
such that $I_\sigma(\rho,\phi, z)=I_\sigma(\rho,\phi+\pi/l, z).$ The total intensity $I_{\textrm{tot}}=\sum\limits_\sigma I_\sigma$ satisfies the same periodicity condition and can be expressed as:
\be
I_{\textrm{tot}}(\rho,\phi, z)=a_{\textrm{tot}}(\rho,z)+b_{\textrm{tot}}(\rho,z)\cos^2(l\phi + c_{\textrm{tot}}(\rho,z)).
\label{eqn:abc2}
\ee

In the ideal paraxial case \cite{Franke-Arnold2007}, the polarization of the electric field is not affected by the focusing, and in general  $a_\sigma = c_\sigma=0$ However, in the nonparaxial regime the latter no longer holds.
The most important parameters of the resulting optical potential are the values of $a_{\textrm{tot}}, b_{\textrm{tot}}, c_{\textrm{tot}}$ in the proximity of intensity maxima, determining the potential depth, the trapping frequencies, and relative shear of the lattice rings.
We find that in general the parameter $c$ is  close to zero, near the focal plane. 

To fully control the relevant energy scales, it is imperative to have independent control of azimuthal lattice depth and radial trapping frequency. To achieve this, an intensity imbalance of the lasers $LG_{l,p}$ and  $LG_{-l,p}$ creating the azimuthal lattice is employed. The resulting potential is again of the form \eqref{eqn:abc}, with $a_{\textrm{tot}} \sim I_l + I_{-l}$ and $b_{\textrm{tot}} \sim \sqrt{I_l I_{-l}}$.  The independent control of the two coefficients,  allows to adjust the radial trapping frequency (depending both on $a_{\textrm{tot}}$ and $b_{\textrm{tot}}$ in (\ref{eqn:abc2})) independent of the azimuthal potential, depending mainly on $b_{\textrm{tot}}.$ In this work we have assumed that the radial trapping frequency of the lattice sites is tuned to make the excitation energy of the radial mode almost equal to the excitation in the $\phi$ direction (see also Fig.~\ref{fig:BlochBand}). 

\subsection{Band Structure calculation}
\label{app:band}

When computing the band structure of the total potential we assume that the axial lattice separates the full potential into a~series of decoupled 2D problems. This assumption is valid close to the focal plane, where the beam is almost parallel to the optical axis. The  resulting Schr\"odinger equation for a~single atom in a~2D potential reads as
\begin{align}
\lr{-\frac{\hbar^2}{2m}\Laplace+V(\rho,\phi)}\Psi(\rho,\phi)=E\Psi(\rho,\phi)
\label{eqn:standard}
\end{align}

In the radial coordinates (energy is measured in units of the recoil energy $E_R = \frac{\hbar^2k^2}{2m}, k=\frac{2\pi}{\lambda},$ the length unit is defined by setting $k=1$) we obtain: 
\be
E\psi = -\left(\frac{1}{\rho}\partial_\rho + \partial_{\rho\rho} \right) - \frac{1}{\rho^2}\frac{\partial^2\psi}{\partial \phi^2} + V(\rho,\phi) \psi.
\label{eqn:angular}
\ee

We want to find the Bloch states  $\psi_q$,
\be
\psi_q\left(\rho,\phi \right)=u_q(\rho,\phi)e^{i\frac{l q}{\pi} \phi},
\ee
where $u$ is a~$\frac{\pi}{l}$ periodic function in the $\phi$ direction.
Expanding $\psi_q,$ in a~Fourier series in the $\phi$ variable we get
\begin{align}
u_{q}(\rho,\phi)=\frac{1}{\sqrt{2l}}\sum_{m=-\infty}^{\infty}e^ {i2m l \phi}d_{q}(\rho,m)
\label{eqn:exp}
\end{align}
where $q$ is the dimensionless quasi-angular momentum.
The 2D Schr\"odinger equation (\ref{eqn:angular}) becomes then an eigenproblem expressed as a~system of coupled differential equations in 1D real space. We solve these equations using sparse matrix diagonalization of a~discretized real space equation. 

The formal singularity of the eigenproblem (\ref{eqn:angular}) at $\rho=0$ is resolved by imposing proper boundary conditions. Indeed, all components of Eq.~(\ref{eqn:exp}) corresponding to a~nonzero angular phase dependence must vanish at the origin, while the only component with no angular phase dependence, $d_0(\rho,0)$ must satisfy: $\left.\partial_\rho d_0(\rho,0) \right|_{\rho=0}=0.$
\begin{figure}
	\includegraphics[width=8cm]{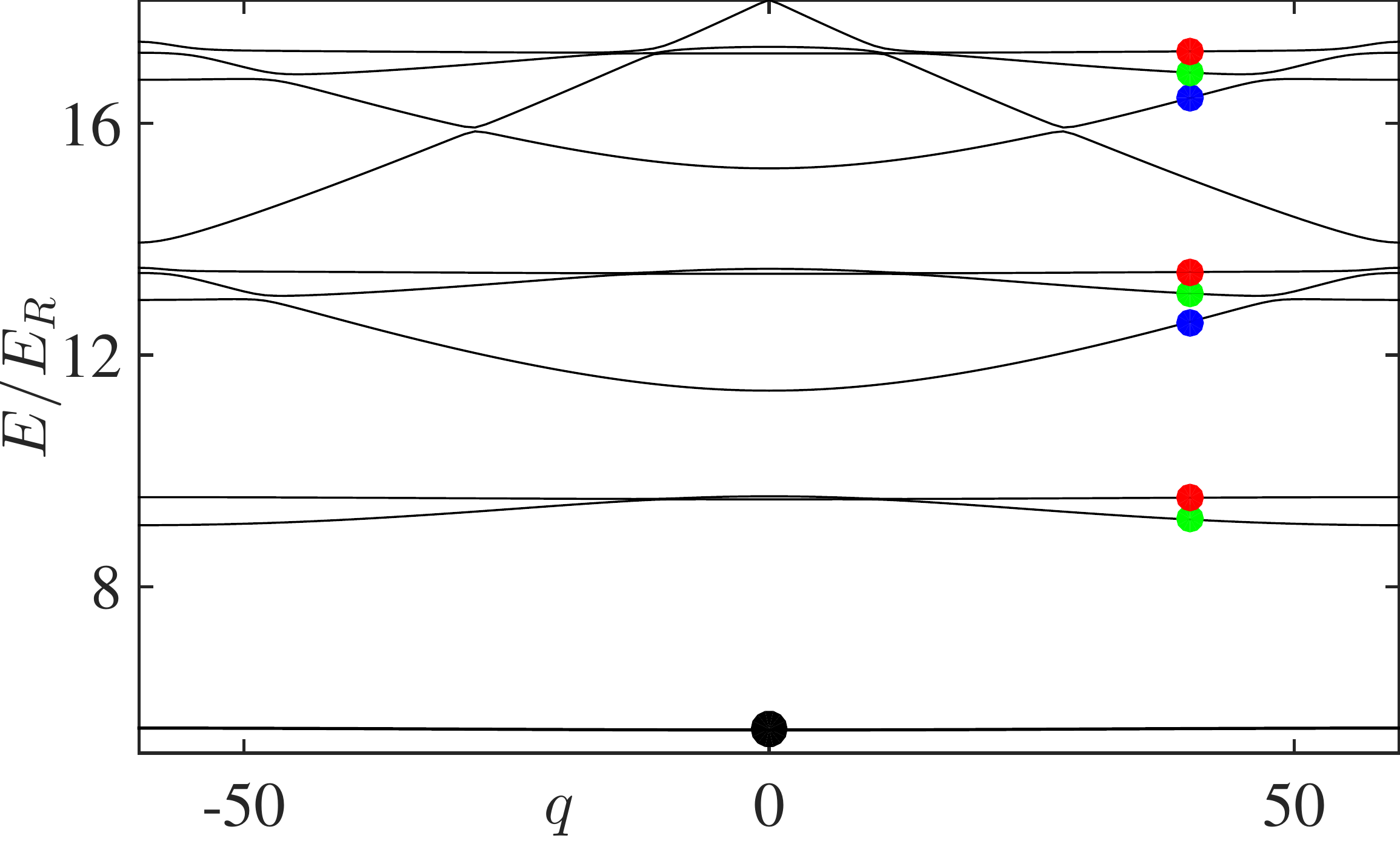}
	\caption{(Color online) Bloch Band structure for a~potential with ring geometry with ring geometry defined in terms of  $\textrm{LG}_{60,15}$ modes. Here $V_z=7 E_R, V_\phi=8.5 E_R.$ The energy is measured relative to energy well depth. The almost-degenerate lowest excited Bloch bands are azimuthally excited (green), radially excited (red). Black dot marks the well-defined lowest Bloch band, while blue dots denote second azimuthally-excited bands. Soft radial modes spawn a~whole ladder of states characterized by azimuthal excitation. Exact formal identification of "radial" and "azimuthal" excitations is not possible due to weak $\rho-\phi$ coupling. }
	\label{fig:BlochBand}
\end{figure}

The numerical solution of the eigenproblem given by Eq.~\ref{eqn:standard} gives rise to the Bloch spectrum shown in Fig.~\ref{fig:BlochBand}. It features a~lowest Bloch band separated from all excited states.  As the problem Eq.~\ref{eqn:standard} is weakly-nonseparable, the overall structure of the excited Bloch states in the spectrum is well described by elementary radial and azimuthal excitations, coupled by very small avoided crossings.

\subsection{Tight-binding and Hubbard parameters}
The purpose of this section is twofold. First, we microscopically compute the hopping integrals $J^\phi_{\mathbf{j}}$ for the cylindrical optical lattice. Second, while the focus of the discussion in the main text is on spinless fermions, we here present the on-site Hubbard interaction parameters of the cylindrical lattice potential. These parameters are relevant if the cylindrical optical lattice is loaded with bosons which is experimentally equally conceivable.

From the numerical solution of Eq.~\ref{eqn:standard} we can extract a~set of Wannier functions in a~standard way \cite{Kohn1959}.The Hubbard parameters describing the contact interaction $U_{\mathbf{j}}=\frac{4\pi a~\hbar^2}{m}\int \textrm{d}^3 r |w_{\mathbf{j}}(\vec r)|^4$ and the hopping integrals $J^\phi_{\mathbf{j}}$ are determined from band computation and construction of the Wannier functions. 

\begin{figure}
	\includegraphics[width=\columnwidth]{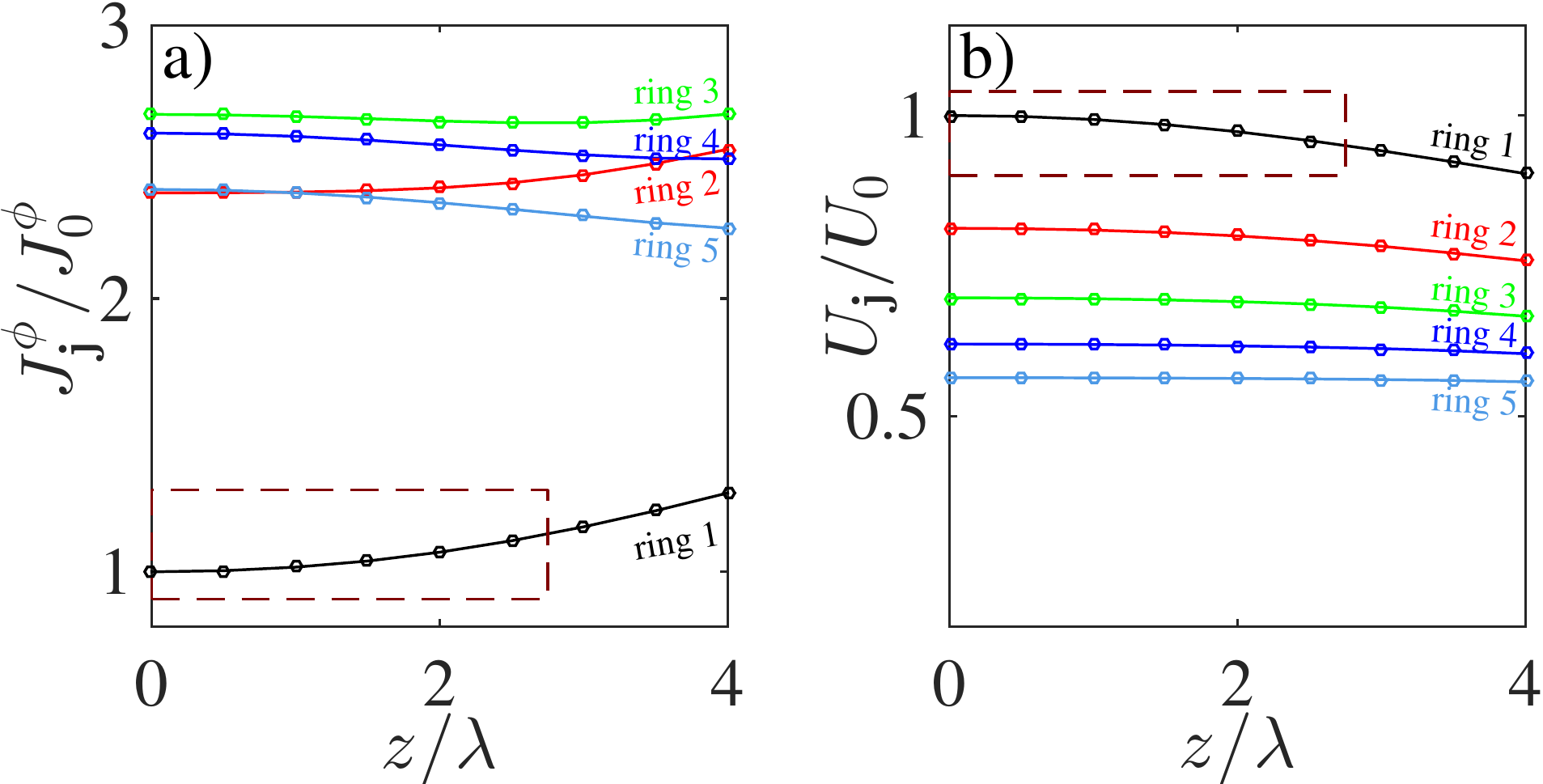}
	\caption{(Color online) Panel a) shows the axial hopping amplitudes $J^\phi_{\mathbf{j}}$ for different rings of the lattice potential. In this setup other hopping directions are prohibited by energy conservation. Panel b) shows the two-body interaction integrals $U_{\mathbf{j}}$ in the respective sites. In both panels the dashed red frame marks lattice sites that are part of the cylinder considered in the main text (sites for $z<0$ are not shown due to symmetry). The potential parameters are: $f/w_0=10.35, V_\phi=8.5, V_z=7 E_R$.}
	\label{fig:UJ}
\end{figure}

The $U_{\mathbf{j}}, J_{\mathbf{j}}^\phi$ parameters are constant within each ring formed by lattice sites due to rotational symmetry.
 
In Fig.~\ref{fig:UJ} we show curves denoting the $U, J$ parameters for different potential rings. Different curves show values for the innermost, second-innermost etc. rings. We note here that the azimuthal lattice depth in the innermost ring is the deepest one, which results in the lowest hopping rates shown in Fig.~\ref{fig:UJ}. For larger ring numbers the increase of radius and the corresponding distance between the lattice sites eventually leads to a decrease of the hopping rate.

\section{Microscopic modelling of artificial magnetic fields}
\subsection{Rotation of the optical lattice}
\label{sec:rotation}

The flux $\Phi_z$ through the cylinder is implemented by rotation of the lattice in the azimuthal direction. A~slight difference of frequencies of lasers creating the azimuthal optical lattice $\omega_1-\omega_2=\Omega,$ causes its rotation around the optical axis with angular velocity $\Omega$ \cite{Franke-Arnold2007}. The time dependent transformation
\begin{align}
U(t)=\exp\left(\frac{i}{\hbar}L_z \Omega t\right)=\exp\lr{-\frac{\Omega t}{l}\frac{d}{d\phi}}
\end{align}
leads  in the co-rotating frame to a~modified Hamiltonian as defined by the Schr\"odinger equation
\be
i\hbar \partial_t {\psi} = \left({H + i\Omega \hbar \frac{\partial}{\partial \phi}} \right){\psi}.
\label{eqn:hamef}
\ee

\begin{figure}
	\includegraphics[width=8cm]{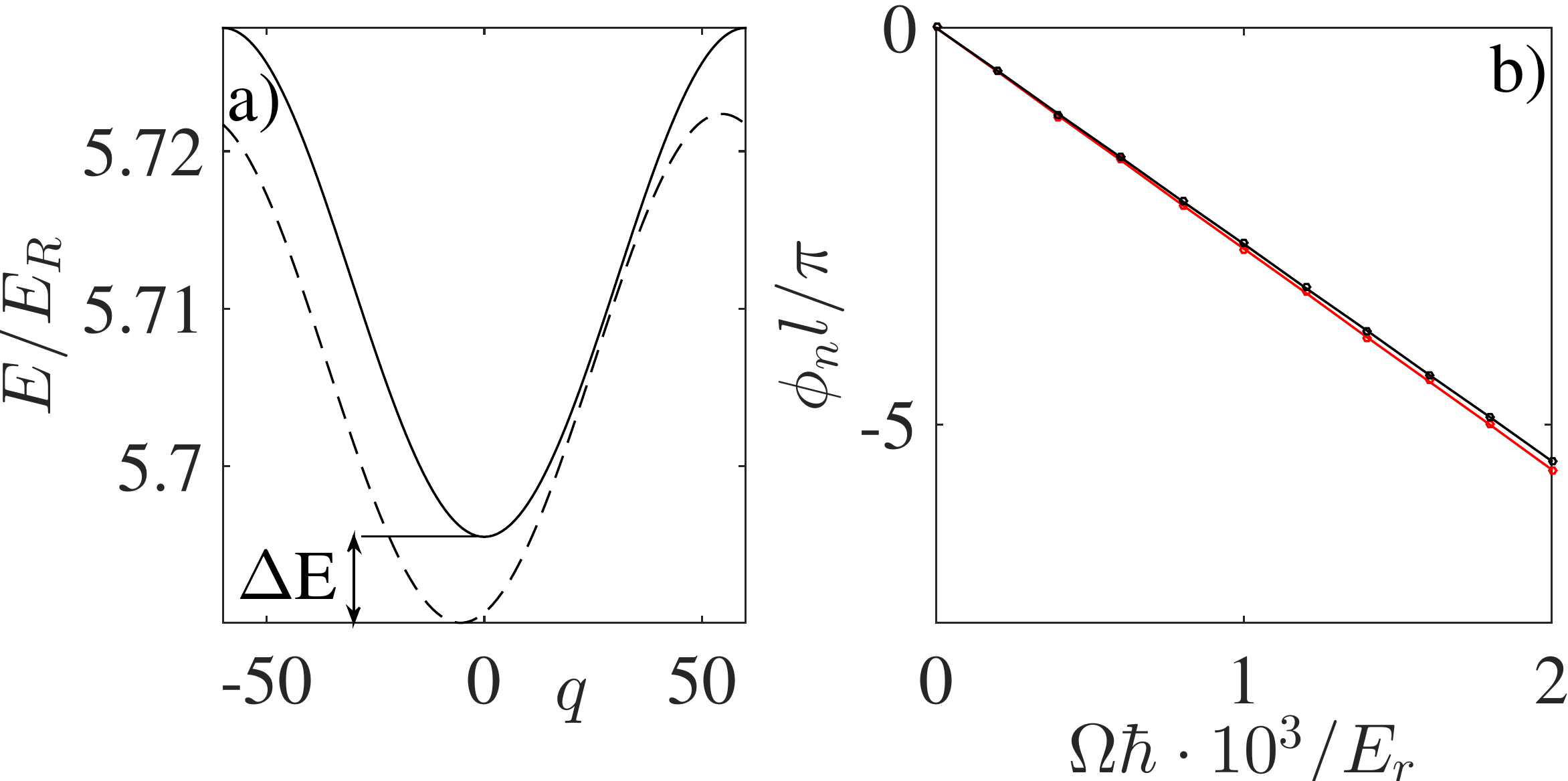}
	\caption{(Color online) Panel a) shows effect of rotation on energy levels of band computation. In addition to the shift of the dispersion relation in the dimensionless quasi angular-momentum $q \to q+2m \Omega r^2/\hbar$ it introduces also an irrelevant offset of the energy $\Delta E$. Panel b) shows dependence of flux  $\Phi_{{z},n}$ threaded by imprinting the phase $\phi_z$ on azimuthal hoppings on the ring number $n$ in cylinder configuration.
	Colors Black/Red indicate fluxes $\Phi_{z,0}, \Phi_{z,5}$ as a~function of $\Omega.$ Rings indicated blue are beyond the range of the cylinder considered in the main text.
	}
	\label{fig:rotationTightBinding}
\end{figure}

At the level of the~tight-binding description, the extra term just amounts to a~shift in the angular quasi-momentum. The corresponding shift of the Bloch bands is shown in Fig.~\ref{fig:rotationTightBinding}a. Formally, this shift is reflected in a~phase imprinted on the azimuthal hopping amplitudes $J^\phi_z \to J^\phi_z e^{i\phi_{z,n}}.$ The phases $\phi_{z,n}$ are slightly different for different potential disks. For the specific set of parameters presented in this work the dependence of $\phi_{z,n}$ on disk number $n$ is presented in Fig \ref{fig:rotationTightBinding}. The rotation corresponds to threading an artificial magnetic field in $z$ direction, and therefore the flux threaded through different rings depends on their radii.

For the simulation of the Laughlin experiment [see Fig.~3 in main text], we consider in this work a~cylinder with an axial length of $11$ sites around the focal plane (with site no. 6 being in the focal plane). In this case, if the phase imprinted on a~lattice ring in the focal plane  is $\Phi_{z,0}=14\pi$, the phase imprinted on the outermost rings (5th nearest neighbours) is $\Phi_{z,5}\approx 1.02 \Phi_{z,0}$. Such a small variation of the probe flux does not have qualitative effects on the Laughlin experiment.

\subsection{Raman-induced hoppings}

A Raman-assisted tunnelling scheme based on a single pair of lasers, requires a linear tilt of the optical lattice in axial direction.
In our setup this requires two steps: 1)~flattening of the natural single particle offsets $E_\mathbf{j},$ 2)~application of a linear gradient by e.g. a magnetic field.

Regarding 1), the natural strong variation of the single-particle on-site energy offsets $E_\mathbf{j}$ can be flattened using external Gaussian beams, as employed to flatten inhomogeneous optical lattices \cite{Schneider2015}. 
The necessary correcting potential  $V_c(\vec r)$ should ideally depend only on the $z$ coordinate, and should vary over the distance of a~few wavelengths.

These properties are realized e.g. by a~potential created by Gaussian beams propagating in the perpendicular direction to the optical axis, focussed at the the same point as the lasers creating the lattice. In general, such a~beam, in the paraxial description, propagating along the $x$ axis, gives rise to the optical potential:
\be
V_c\sim |\vec E|^2 \sim \frac{| E_0|^2 }{w_y(x) w_z(x)} \exp\left(-2 \left(\frac{z^2}{w_z(x)^2} + \frac{y^2}{w_y(x)^2}\right)\right).
\label{eqn:nine}
\ee
with $w_i(x)=w_0\sqrt{1+(x/x_R)^2}, x_R={w_0^2 \pi}/{\lambda}.$
To minimize the residual variance of $E_\mathbf{j}+V_c$ on, one can focus it by a~cylindric lens in the $x-y$ plane, i.e. $w_y^0 \to \infty.$ 

An alternative energy correction scheme could be based on strongly focusing correction fields (again of Laguerre-Gaussian type) using the same optical setup as for the lattice lasers. To achieve maximal dependence with the $z$ coordinate, a~beam with $p=0$ should be used. This approach retains perfect rotational invariance of the correction field on the innermost ring of the lattice.

In what follows we have simulated an exemplary Gaussian correction scheme (the same parameters as those used in Fig. 3. in the main text) [Eq.~(\ref{eqn:nine})] . The optimization of parameters of correction lasers is performed as follows. First, by a tight-binding computation, we determine the single-particle energy levels $E_\mathbf{j}$ of the cylindric potential as a~function of ring number. As a correction potential, we use two Gaussian potentials, one with paraxial waist $w_1,$ the other with $w_2.$  Then we define the following auxiliary function with free parameters $\alpha, \beta, w_1, w_2$:  
\bea
&&F(\alpha,\beta,w_0,w_1)=\nonumber\\
&&\!=\sum\limits_{i=L}^{i=R-1} \left| c_{i+1}-c_i\right|^2-\gamma | E_{R+1}-E_R-\delta |,\\
\!\!\!\!\!\!\!\!\!\!&&c_i=E_i-\!\alpha V_c(w_1,i)- \beta V_c(w_2,i)
\eea
and Lagrange multiplier $\gamma.$ Minimization of $F$ ensures that $E_i$ are flat between sites $L=-5$ and $R=5$, and that the energy $\delta$ barriers defining boundary condition are high, allowing a hard wall description. Here we fix $\delta$ at $\delta \approx 5.5 J^z.$ The energies of single-particle states of particular rings are then almost equal $E_i \approx E + \Delta E_i$ with $|\Delta E_i | \leq 0.08 J^z.$ 

\begin{figure}
\includegraphics[width=8.6cm]{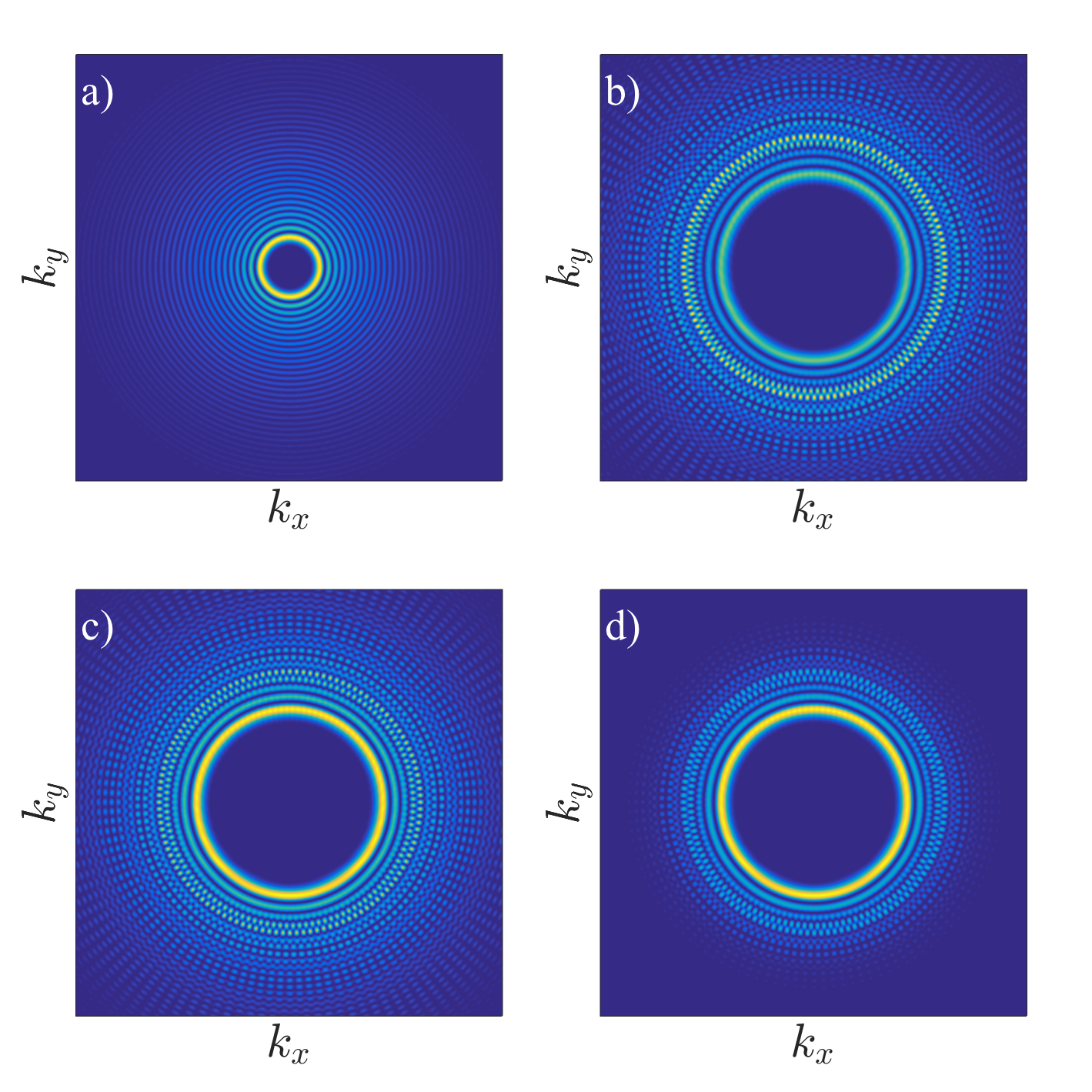}
\caption{(Color online) Time-of-flight images of a~lowest band Bloch function in a~cylindrical optical potential, released abruptly from the innermost ring of a~potential in the focal plane. Here $V_z=7 E_R, V_\phi=8.5 E_R, f/w_0=10.35.$ Panels show images of different quasi-angular momenta $q.$ Subsequent panels show:  a) $q=15,$  b) $q=50.$ Remaining panels  show $q=50$ as well in the case when the optical potential is switched off adiabatically, to allow for adiabatic band mapping \cite{Bloch2008a}. Panel c) deals with the case when adiabatic band mapping is performed by ramping down just one of laser creating the lattice, thus reducing $V_\phi,$ from $8.5 E_R$ to a final value $0.5 E_R,$ while in panel d) intensity of both lattice lasers is homogeneously decreased to the point when $V_\phi=0.5 E_R.$
}
	\label{fig:tof}
\end{figure}

{\emph{Microscopic Raman-assisted hopping parameters}}. - With the application of a linear tilt of the onsite energies of $\Delta$ per lattice constant the bare tunnelling elements in axial direction are off-resonant, $J^{z,\textrm{tun}}_\mathbf{j}\ll \Delta$. Tunnelling can then be restored via a two photon Raman process. This allows also to introduce the flux $\Phi_r$. As discussed in the main text, to create $\Phi_r$, requires a~pair of counter-propagating Laguerre-Gaussian beams. The Raman-induced tunnelling amplitude between two $s$-band states is given by $J^z_\mathbf{j} \approx J^{z,\textrm{tun}}_\mathbf{j} J_1(2\frac{\tilde{\Omega}_\mathbf{j}}{\Delta} \eta_z \eta_{\phi})$  \cite{Ketterle2013}, where $\Omega$ denotes the height of the time-dependent potential created by the Raman lasers. The quantities $\eta_z = \langle w_{\mathbf{j}} | \exp(2 i k z) | w_{\mathbf{j}+\hat{z}}\rangle$ and $\eta_\phi = \langle w_{\mathbf{j}} | \exp(\frac{2}{3}l \phi  ) | w_{\mathbf{j}+\hat{\phi}}\rangle$ are defined via integrals involving the Wannier functions $w_\mathbf{j}$.
For our setup, $\eta_\phi$ and $\tilde{\Omega}_{\mathbf{j}}$ are $\mathbf{j}$-dependent due to focussing-induced variation of the azimuthal lattice depth. However, the variance of $\eta_\phi$ is small, and the Raman lasers, operated far from the diffraction limit, exhibit only a minor intensity variation over the innermost rings. In conclusion, our microscopic calculations show that the resulting variation in the Raman-induced hopping is less than 1\% over the whole cylinder and is thus much smaller than the variation in the natural azimuthal hopping.

\section{Time of flight images}

Time-of-flight (TOF) imaging  is a~standard experimental method to probe a~single particle Green's function $G(r,r')=\langle \hat\psi(r,t)\hat\psi^\dagger(r',t)\rangle$ (see \cite{Bloch2008a}). After the atomic cloud is abruptly released from the confinement and allowed to expand freely, an absorption imaging is performed along the optical axis $z$ which measures the density of the gas after time $t_{TOF}$. The measured density is the full 3D density of the cloud, i.e.,  $n(\mathbf{x})=\left(\frac{m}{\hbar t_{TOF}}\right)^3 |\tilde{w}(\mathbf{k})|^2 {\cal{G}}(\mathbf{k}=\frac{m\mathbf{x}}{\hbar t_{TOF}})$, integrated along $z$ direction, where $w$ is the Wannier state of the occupied Bloch band.

Fig.~\ref{fig:tof} shows the absorption image for a~single Bloch state calculated using the microscopic description of the potential modelled in this article (with $V_z=7E_R, V_\phi=8.5E_R, f/w_0=10.35$), and a~fixed quasi-angular momentum $q$. It has the structure of a~ring with a~radius proportional to $|q|.$ This allows to measure the distribution of the absolute value of the angular quasi-momentum. Rotating the lattice as described in \ref{sec:rotation} allows to shift the quasimomentum distribution and measure it as a~function of $|q-q_0|$ with the experimentally controllable offset $q_0$ .

An alternative approach is to moderate the speed with which the lattice is switched off before the free expansion phase \cite{Bloch2008a}. This enables the Bloch states to adiabatically follow the decreasing potential height. This converts the initial quasi-angular momentum for the deep lattice to the ordinary angular momentum distribution. This allows to reduce higher order peaks for large angular quasi-momenta $q$ as seen in Fig.~\ref{fig:tof}b).

We consider two cases of adiabatic ramping down of the potential. In the first case [see Fig.~\ref{fig:tof}c)] only one of the lasers creating the azimuthal lattice is ramped down (and the other is slightly ramped up to keep the radial curvature of sites constant). In terms of numbers the initial azimuthal lattice depth $V_\phi=8.5E_R$  is reduced to $V_\phi=0.5$. In the second case [see Fig.~\ref{fig:tof}d)] both lasers creating the azimuthal lattice are ramped down until the depth of the azimuthal lattice reaches a final value of $V_\phi=0.5$. Afterwards the lattice is switched off abruptly (close to $V_z\approx 0$ , adiabaticity is not possible). The Lasers creating the axial lattice are also switched off in both scenarios. With this technique the best discrimination between different $|q|$ can be obtained in the TOF image.

\end{document}